\documentclass[11pt,a4paper]{article}
\pdfoutput=1
\usepackage{
	jheppub,
	amssymb,
	amsthm,
	amscd,
	amsfonts,
	bbold,
	MnSymbol,
	slashed
}
\usepackage[normalem]{ulem}
\usepackage{tocloft}

\def\a{\alpha}
\def\b{\beta}
\def\g{\gamma}
\def\d{\delta}
\def\e{\epsilon}
\def\h{\eta}
\def\q{\theta}
\def\c{\chi}
\def\p{\pi}
\def\f{\varphi}
\def\r{\rho}
\def\l{\lambda}

\def\m{\mu}
\def\n{\nu}
\def\y{\psi}
\def\D{\Delta} 
\def\eh{\hat{\epsilon}}
\def\ah{\hat{\alpha}}
\def\bh{\hat{\beta}}
\def\tK{\tilde{K}}
\def\sF{\slashed{F}}

\def\mc{\mathcal}
\def\mH{\mathcal{H}}

\def\tx{\tilde{x}}
\def\XX{\mathbb{X}}
\def\CP{\mathbb{CP}}
\def\RR{\mathbb{R}}
\def\SS{\mathbb{S}}
\def\L{\Lambda}
\def\G{\Gamma} 
\def\ff{\phi}
\def\dt{\partial}
\def\fr{\frac}

\def\tdt{\tilde{\partial}}
\def\mE{\mathcal{E}}
\def\mM{\mathcal{M}}
\def\mN{\mathcal{N}}
\def\mA{\mathcal{A}}
\def\mB{\mathcal{B}}
\def\mC{\mathcal{C}}

\title{Non-abelian fermionic T-duality in supergravity}

\author[a,b,c]{Lev Astrakhantsev,}
\author[a,d]{Ilya Bakhmatov,}
\author[b,d]{Edvard T. Musaev,}

\affiliation[a]{Institute for Theoretical and Mathematical Physics, Lomonosov Moscow State University, \\Lomonosovsky avenue, Moscow, 119991, Russia}
\affiliation[b]{Moscow Institute of Physics and Technology, Institutskii per. 9, Dolgoprudny, 141700, Russia}
\affiliation[c]{Institute of Theoretical and Experimental Physics, B. Cheremushkinskaya, 25, 117218,\\Moscow, Russia}
\affiliation[d]{Kazan Federal University, Institute of Physics, Kremlevskaya 16a, Kazan, 420111, Russia}

\emailAdd{lev.astrakhantsev@phystech.edu}
\emailAdd{ibakhmatov@itmp.msu.ru}
\emailAdd{musaev.et@phystech.edu}

\abstract{Field transformation rules of the standard fermionic T-duality require fermionic isometries to anticommute, which leads to complexification of the Killing spinors and results in complex valued dual backgrounds. We generalize the field transformations to the setting with non-anticommuting fermionic isometries and show that the resulting backgrounds are solutions of double field theory. Explicit examples of non-abelian fermionic T-dualities that produce real backgrounds are given. Some of our examples can be bosonic T-dualized into usual supergravity solutions, while the others are genuinely non-geometric. Comparison with alternative treatment based on sigma models on supercosets shows consistency.}


\arxivnumber{2101.08206}

\begin{document}
\maketitle

\section{Introduction}

A very fruitful approach to the analysis of the structure of physical theories is that based on their symmetries. It allows to overcome difficulties related to a possible bad choice of the degrees of freedom and to search for a better one. A textbook example is the reformulation of the Maxwell theory in terms of four-dimensional tensors rather than three-dimensional field strength vectors, which makes manifest the Lorentz symmetry of the Maxwell equations. String theory and supergravity possess a wealth of duality symmetries, which relate background field configurations that are equivalent from the point of view of the string. Restricting the narrative to perturbative dualities of the $d=10$ superstring sigma model, one recalls that it enjoys bosonic T-duality (abelian, non-abelian and more generally Poisson-Lie) and fermionic T-duality symmetries. The standard abelian bosonic T-duality transformation starts with the string in a background that has $d$ commuting Killing vectors, representing the isometry group U(1)$^{d}$. Gauging each symmetry and introducing $d$ Lagrange multipliers to preserve the amount of the worldsheet degrees of freedom, we can rewrite the model in the first order formalism. Integrating out the Lagrange multipliers we recover the initial theory, while integrating out the corresponding gauge fields leads to the same sigma model, however on a different background~\cite{Buscher:1987qj,Buscher:1987sk}. The two backgrounds are related by the so called Buscher rules, which in particular mix the metric and the 2-form $b$-field degrees of freedom~\cite{Buscher:1985kb,Alvarez:1993qi}. The Buscher procedure can be generalized to non-abelian isometry groups, in which case the symmetry is referred to as non-abelian T-duality~\cite{delaOssa:1992vc} and Poisson-Lie T-duality~\cite{Klimcik:1995ux,Klimcik:1995jn}. For more details on T-duality symmetry and its global structure see e.g.~\cite{Giveon:1994fu,Alvarez:1994dn,Cavalcanti:2011wu,Bugden:2019wnc,Demulder:2019bha}.

An extension of this idea to the superspace setting, while conceptually straightforward, was not developed until much later~\cite{Berkovits:2008ic}. Instead of a Killing vector isometry one assumes invariance of the background superfields under a shift of a fermionic coordinate in superspace. This implies the existence of an unbroken supersymmetry, parameterized by a Killing spinor field. Starting from the superstring action in a manifestly spacetime supersymmetric formalism such as Green-Schwarz or Berkovits pure spinor sigma model, fermionic version of the Buscher procedure yields the transformation rules for the supergravity component fields~\cite{Berkovits:2008ic}. These rules only affect the dilaton and the field strength fields from the Ramond-Ramond sector, without changing the initial values of the metric and the NS-NS 2-form. Originally introduced as a building block in the $AdS_5 \times S^5$ T-self-duality scheme, fermionic T-duality is a component in the string theory description of the amplitude/Wilson loop correspondence and the dual superconformal invariance of super-Yang-Mills scattering amplitudes~\cite{Alday:2007hr,Alday:2008yw,Beisert:2008iq,Berkovits:2008ic}. For a review of fermionic T-duality and some related developments see~\cite{OColgain:2012si}.

Results of \cite{Adam:2009kt} on self-duality of integrable Green-Schwarz sigma models under fermionic T-duality suggest, that such self-duality could imply integrability~\cite{Roiban:2010kk,Drummond:2010km,Alday:2010kn,AntonioPittelli:2016vua,Tarrant:2017sfq}. In particular this is true for sigma models on supercosets based on AdS${}_p\times \SS^p$, for $p=2,3,5$, which are both fermionic T-self-dual and integrable~\cite{Abbott:2015ava,Abbott:2015mla}.   Robust support for this point could come from the  AdS${}_4\times \CP^3$ sigma model, which is known to be integrable~\cite{Sorokin:2010wn, Cagnazzo:2011at} (for a review see~\cite{Sorokin:2011mj}). However, despite many attempts, T-self-duality of AdS${}_4\times \CP^3$ has not been shown to the moment~\cite{Bakhmatov:2010fp,Colgain:2016gdj}. Part of the reason is that the duality transformation is highly restricted by the commutativity constraints. Effectively these restrict the T-duality to complexified fermionic directions in superspace, which leads to complexified supergravity backgrounds. As a matter of fact, even a single fermionic T-duality with respect to a Majorana Killing spinor of type II $d=10$ supergravity is non-abelian by default, in the sense that the supersymmetry generator does not have a vanishing bracket with itself $\{Q^a, \bar{Q}_b\} = (\G^m)^a{}_b P_m$. Complexifying the Killing spinor allows to satisfy the abelian constraint and thus make fermionic T-duality consistent. In order to end up with a real background one usually performs a chain of fermionic T-dualities, arranged in such a way that the imaginary components of fields generated in the process cancel out~\cite{Berkovits:2008ic,Bakhmatov:2009be,Bakhmatov:2011aa}. 

A natural goal is to modify or extend the abelian fermionic T-duality procedure in such a way, that real backgrounds would be easier to access. A generalization of fermionic T-duality transformation was proposed by considering the most general fermionic symmetry of type II supergravity~\cite{Godazgar:2010ph}. It was shown that the transformation can potentially give real backgrounds, however, to our knowledge, no examples have been presented so far. An alternative approach, which we pursue in this work, is to relax the abelian constraint and develop the non-abelian fermionic T-duality. This makes using real (Majorana) Killing spinors possible.

However, this comes at a price. As we will see, a natural arena for the non-abelian fermionic T-dual backgrounds appears to be double field theory (DFT). The backgrounds that we get are in general non-geometric, in the sense that they depend on both the original and the T-dual spacetime coordinates. This is to be expected, because non-abelian fermionic isometries do not form a sub-algebra within the isometry super-algebra: they close on a bosonic generator. Thus, dualizing the non-abelian fermionic isometries alone, we find ourselves in double field theory rather than the standard supergravity. The DFT solution can then be taken further along the T-duality orbit by a bosonic duality with respect to the remaining element of the sub-algebra. A surprising result is that this orbit is not always geometric, i.e.\ completing the sub-algebra with a bosonic T-duality may still leave us in the realm of DFT.

The superstring dynamics on DFT backgrounds is well-defined in the same sense in which it is well-defined on backgrounds of the generalized supergravity~\cite{Arutyunov:2015mqj,Fernandez-Melgarejo:2018wpg}. The latter are known to form a subset among solutions to DFT equations of motion. As an illustration of this statement we compare our results with the recent unified treatment of bosonic and fermionic T-dualities that appeared in~\cite{Borsato:2018idb}. The authors show explicitly that the Green-Schwarz superstring on a coset superspace can be T-dualized along both bosonic and fermionic isometries. The known non-abelian bosonic T-duality transformation rules are re-derived, and the conditions are mentioned under which the fermionic transformations reproduce the known formulae of abelian fermionic T-duality. In section \ref{sec:sigma} we compare our results to those of~\cite{Borsato:2018idb} and observe that for the particular case of supercoset backgrounds these are consistent, and hence the non-abelian fermionic T-duality transformation rules that we analyse indeed keep the sigma model on a supercoset invariant.

We start with a brief recap of abelian fermionic T-duality in section~\ref{sec:ferm}, then describe our proposed non-abelian extension. This is followed by a concise review of double field theory in section~\ref{sec:dft}. We check that the dilaton field equation is satisfied by a non-abelian fermionic T-dual background. Then in section~\ref{sec:sigma} we make connection with the non-abelian duality transformation that was derived for generic supercosets. Several explicit examples of non-abelian fermionic T-duals follow in section~\ref{sec:examples}, and we formulate some conclusions in section~\ref{sec:conc}.

\section{Fermionic T-duality}\label{sec:ferm}

Fermionic T-duality requires invariance of the background superfields under the shift isometry of some fermionic direction. Such shift symmetry is equivalent to an unbroken supersymmetry, which in type II supergravity is defined by a pair of Killing spinors~$\e,\eh$. Depending on whether we are in Type IIA or Type IIB theory these would be of the opposite or the same chirality respectively. To avoid confusions it is important to mention that  $\e,\eh$ is a pair of Killing spinors that fix a \emph{single} fermionic direction in the $\mathcal{N}=(1,1)$ or $\mc{N}=(2,0)$ $d=10$ superspace. Anticommutation constraint for the pair is given by vanishing of the Killing vector field
\begin{equation}\label{constr}
    \tK^m = \left\{\,
    \begin{aligned}
         &\e \bar\g^m \e - \eh \g^m \eh \quad \mathrm{(IIA)}\\
          &\e \bar\g^m \e + \eh \bar\g^m \eh \quad \mathrm{(IIB)}
    \end{aligned}
    \,\right\} \overset{!}{=} 0 \qquad \mathrm{abelian~constraint.}
\end{equation}
Our spinor and gamma matrix conventions are summarized in the Appendix \ref{app:spinor}. As demonstrated there, $\tK^m$ is simply proportional to a commutator of the supersymmetry transformation with itself, $[\d_{\e,\eh},\d_{\e,\eh}]$. It is important to observe that in a Majorana basis where $\g^0 = -\bar\g^0 = 1$, $\tK^0$ is a simple sum of squares of all components of a spinor. Thus the abelian constraint cannot be satisfied by the standard Killing spinors of type II supergravity, which are real in this representation. As a result, abelian constraint for fermionic T-duality necessitates complexification of the Killing spinors.

The resulting abelian fermionic T-dual background can be deduced via the fermionic version of the Buscher procedure\footnote{Despite its name, fermionic T-duality does not affect the background fermionic fields. Note also that the only NSNS field affected is the dilaton; in particular, the metric is invariant.}:
\begin{equation}
 \begin{aligned}
 \label{fTtrans}
   e^{\ff'}F'& = e^\ff F + 16\,i\, \frac{\e \otimes \eh}{C},\\
   \ff'&=\ff+\fr{1}{2}\log C.
 \end{aligned}
\end{equation}
The background fields that undergo the transformation are the RR bispinor\footnote{E.g.\ in type IIB, $F^{\a\hat\b} = (\g^m)^{\a\hat\b} F_m+ \fr{1}{3!} (\g^{m_1\overline{m}_2m_3})^{\a\hat\b} F_{m_1m_2m_3}+ \fr12\fr{1}{5!} (\g^{m_1\ldots m_5})^{\a\hat\b} F_{m_1\ldots m_5}$.} $F$ and the dilaton $\phi$, while the scalar parameter $C$ is defined by the system of PDEs in terms of the Killing spinors~\cite{Berkovits:2008ic}:
\begin{equation}\label{Cdef}
    \dt_m C = \left\{\,
    \begin{aligned}
          &i(\e \bar\g_m \e + \eh \g_m \eh) \quad \mathrm{(IIA)},\\
         &i(\e \bar\g_m \e - \eh \bar\g_m \eh) \quad \mathrm{(IIB)} .
    \end{aligned}
    \right.
\end{equation}
We denote $\dt_m C = i K_m$ because the above expressions are very similar to $\tilde K^m$. 

In what follows we will see how this transformation can be modified to include the non-abelian case.

\subsection{Non-abelian extension}
\label{sec:natfd}

So what happens when the supersymmetry transformation violates the abelian constraint? Focusing on Type IIB case for definiteness, we have that both $K_m = \e \bar\g_m \e - \eh \bar\g_m \eh$ and $\tK^m = \e \bar\g^m \e + \eh \bar\g^m \eh$ are nonzero. If the field transformation~\eqref{fTtrans} is formally applied in this case, it would not map a supergravity solution to a solution. Let us however take a closer look at $K$ and $\tK$. It is easy to see that they are orthogonal, $\tilde{K}^m K_m=0$, by invoking the Fierz identities for the chiral $d=10$ spinors $\e$ and $\eh$. Since $\tK^m$ is a Killing vector, this implies that $K_m$ can indeed be represented by a derivative of a scalar as in~\eqref{Cdef}, up to terms that vanish identically upon contraction with $\tK^m$. Moreover, Killing spinor equations can be employed to check that $\tK^m$ is divergence free:
\begin{equation}
\begin{aligned}\label{nablaK}
\nabla_m \tilde{K}^m = 2 \e \bar\g^m \nabla_m \e + 2\eh \bar\g^m \nabla_m \eh &=  \e \bar\g^m \left[ \fr12 \slashed{H}_m \e + \frac{e^\phi}{4} \left( \sF_{(1)} + \sF_{(3)} + \fr12\sF_{(5)} \right) \bar\g_m\eh \right] \\ & - \eh \bar\g^m  \left[ \fr12 \slashed{H}_m \eh + \frac{e^\phi}{4} \left( \sF_{(1)} - \sF_{(3)} + \fr12\sF_{(5)} \right)  \bar\g_m\e \right]= 0.
\end{aligned}
\end{equation}
One has to take into account that $\e \slashed{H} \e =0$, $\bar\g^m \sF_{(1)}\bar\g_m=-8 \sF_{(1)}$, $\bar\g^m \sF_{(3)}\bar\g_m=-4 \sF_{(3)}$, $\bar\g^m \sF_{(5)}\bar\g_m=0$ due to the gamma matrix algebra and symmetry properties of the gamma matrices. For our supersymmetry and spinor conventions see appendix~\ref{app:spinor}. 

These observations suggest that the non-abelian fermionic T-dual background can be defined using the same transformation rules~\eqref{fTtrans}, but with the modified prescription for the scalar parameter $C$:
\begin{equation}
    \label{eq:naftd}
    \begin{aligned}
        \dt_m C & = i K_m - i b_{mn} \tK^n,\\
        \tdt^m C & = i \tK^m,
    \end{aligned}
\end{equation}
where $\tdt^m$ denotes derivative with respect to the dual coordinates $\tx_m$ of double field theory, and the $b_{mn}$ term is added in order to make the two equations consistent. Indeed, as we briefly review in the next section, consistency of the double field theory formulation requires the doubled coordinates $\XX^M=(x^m,\tx_m)$ dependence of all fields to comply with the section constraint. For the field $C$, which is the only function of dual coordinates here, the weak and the strong section constraints read
\begin{equation}
\begin{aligned}
\dt_m C \tdt^m C=0, && \dt_m \tdt^m C=0.
\end{aligned}
\end{equation}
One can immediately notice, that the former is satisfied due to the identity $K_m \tK^m \equiv 0$. The latter can be rewritten as follows
\begin{equation}\label{strong}
   \dt_m \tdt^m C= i \dt_m \tK^m = i\nabla_m \tilde{K}^m - \frac{i \tilde K^m \dt_m g}{2g},
\end{equation}
where the first term on the right hand side vanishes because $\tilde K^m$ is a Killing vector, and $g=\det g_{mn}$. The last term can be turned to zero by choosing adapted coordinates, where $g$ is independent of the direction of the isometry that is given by the vector field $\tK^m$. It is tempting to include the factor $g^{1/2}$ into the definition of  $\tdt^m C$ to make the above hold for any coordinate choice. However, as we show further, explicit examples and comparison with the sigma model approach for coset spaces requires the definition as in \eqref{eq:naftd}. Note, that this does not cause problems with the section constraint of DFT as the transformation \eqref{eq:naftd} is defined only for backgrounds with specific fermionic and coordinate isometries. 

To summarize, our proposed non-abelian fermionic T-duality prescription results in double field theory backgrounds, where explicit dependence on dual coordinate enters via the second equation in~\eqref{eq:naftd}. This dependence can often be removed by a bosonic T-duality with respect to the isometry given by the supersymmetric Killing vector $\tilde K^m$, which then leaves us with a proper supergravity solution.

The modification of the definition of $C$ by the extra $\tilde K^m$ terms that we propose is rooted in the analysis of the DFT constraints and equations of motion. We do not attempt a first principles derivation of a non-abelian Buscher procedure in the DFT setting and solvinig the supergravity constraints in the superfield formalism for DFT~\cite{Hatsuda:2014qqa,Bandos:2015cha,Bandos:2016jez,Cederwall:2016ukd}, leaving that for a separate work. In the following section we do check that the fermionic non-abelian T-duals defined by~\eqref{fTtrans},~\eqref{eq:naftd} are actually solutions of DFT.

\section{Double field theory and equations of motion}\label{sec:dft}

We will provide a short overview of the necessary concepts of double field theory, which enables us to quickly get to the equations of motion. For more detailed description of the construction see the original works \cite{Siegel:1993th,Siegel:1993bj,Hohm:2010jy,Hohm:2010pp,Hohm:2010xe,Hohm:2011nu,Hohm:2011dv,Geissbuhler:2013uka} and reviews \cite{Berman:2013eva,Hohm:2013bwa,Aldazabal:2013sca}.

\subsection{Overview of the DFT formalism}

Double field theory is an approach to supergravity that makes T-duality manifest at the level of the action by doubling the spacetime coordinates. It introduces the usual ‘momentum’ (spacetime) coordinates $x^{m}$ together with new ‘winding’ coordinates $\tilde{x}_{m}$ combined into $\XX^{M}=(x^{m},\tilde{x}_{m})$ and also the covariant constraint
\begin{equation}
    \begin{aligned}
     \h^{MN}\dt_M \bullet \dt_N \bullet = 0, && 
     \h^{MN}=
        \begin{bmatrix}
            0 & \d_m{}^n \\
            \d_n{}^m & 0
        \end{bmatrix}.
    \end{aligned}
\end{equation}
The condition, called the section constraint, effectively eliminates half of the coordinates and ensures closure of the algebra of local coordinate transformations~\cite{Berman:2012vc}.

The action of ten-dimensional supergravity on such doubled space can be made manifestly covariant under the global O$(10,10;\RR)$ T-duality rotations as well as the local generalized diffeomorphisms, which include standard diffeomorphisms, gauge transformations of the Kalb-Ramond $b$-field, and transformations exchanging momentum and winding coordinates. The action is background independent and takes the form
\begin{equation}
\label{act}
S=S_{NSNS}+S_{RR}=\int d^{10}x\, d^{10}\tilde{x} \left( e^{-2d} \mathcal{R} (\mathcal{H},d) + \frac14 (\slashed{\partial}\chi)^{\dagger} S \,\slashed{\partial}\chi \right),
\end{equation}
where the NSNS degrees of freedom are encoded by the invariant dilaton $d$ and the generalized metric $\mH_{MN}$ with its spin representative $S\in \mathrm{Spin}(10,10)$, while the RR field strengths are contained in the spinorial variable~$\c$.

Let us start with the NSNS fields. The invariant dilaton $d$ is simply
\begin{equation}
    d = \phi - \fr14 \log g,
\end{equation}
where $g=\det g_{mn}$. The generalized metric of DFT is an element of the coset space $\mathrm{O}(10,10)/\mathrm{O}(10)\times \mathrm{O}(10)$ and in terms of the background fields is defined as follows
\begin{equation}
 \mathcal{H}_{MN}=\begin{bmatrix} g_{mn}-b_{mp}g^{pq}b_{qn}&b_{mp}g^{pl}\\
-g^{kp}b_{pn}&g^{kl}\end{bmatrix}.
\end{equation}
Varying the action \eqref{act} with respect to the dilaton field $d$ and choosing the representation of $\mathcal{R}$ so that it resembles the dilaton equation of the usual supergravity, we obtain the first equation of motion:
\begin{equation}
\begin{aligned}
\mathcal{R}(\mathcal{H},d) &\equiv  4 \mathcal{H}^{M N} \partial_{M} \partial_{N} d-\partial_{M} \partial_{N} \mathcal{H}^{M N}-4 \mathcal{H}^{M N} \partial_{M} d \partial_{N} d+4 \partial_{M} \mathcal{H}^{M N} \partial_{N} d \\
&+\frac{1}{8} \mathcal{H}^{M N} \partial_{M} \mathcal{H}^{K L} \partial_{N} \mathcal{H}_{K L}-\frac{1}{2} \mathcal{H}^{M N} \partial_{M} \mathcal{H}^{K L} \partial_{K} \mathcal{H}_{N L}=0.
\end{aligned}
\end{equation}
This is the equation that we will later check for specific non-abelian fermionic T-dual backgrounds to ensure that they are DFT solutions. The above equation does not contain RR fields as they do not couple to the dilaton in the DFT action. 

Next consider the variation with respect to the generalized metric $\mathcal{H}_{MN}$. For the NSNS part of resulting field equation one obtains the following expression:
\begin{equation}
\delta S_{NSNS}=\int d^{10}x\, d^{10}\tilde{x}\, e^{-2 d} \delta \mathcal{H}^{M N} \mathcal{R}_{M N}.   
\end{equation}
Here $\mc{R}_{MN}$ is the DFT analogue of Ricci curvature~\cite{Hohm:2011si} and reads
\begin{equation}
\mathcal{R}_{M N} \equiv \frac{1}{4}\left(\delta_{M}^{P}-\mathcal{H}_{M}^{P}\right) \mathcal{K}_{P Q}\left(\delta_{N}^{Q}+\mathcal{H}_{N}^{Q}\right)+\frac{1}{4}\left(\delta_{M}^{P}+\mathcal{H}_{M}^{P}\right) \mathcal{K}_{P Q}\left(\delta_{N}^{Q}-\mathcal{H}_{N}^{Q}\right),
\end{equation}{}
with
\begin{equation}
    \begin{aligned}
    \mathcal{K}_{M N} &\equiv  \frac{1}{8} \partial_{M} \mathcal{H}^{K L} \partial_{N} \mathcal{H}_{K L}-\frac{1}{4}\left(\partial_{L}-2\left(\partial_{L} d\right)\right)\left(\mathcal{H}^{L K} \partial_{K} \mathcal{H}_{M N}\right)+2 \partial_{M} \partial_{N} d \\ &-\frac{1}{2} \partial_{(M} \mathcal{H}^{K L} \partial_{L} \mathcal{H}_{N) K}+\frac{1}{2}\left(\partial_{L}-2\left(\partial_{L} d\right)\right)\left(\mathcal{H}^{K L} \partial_{(M} \mathcal{H}_{N) K}+\mathcal{H}_{(M}^{K} \partial_{K} \mathcal{H}_{N)}^{L}\right). \end{aligned}
\end{equation}
In contrast to the dilaton equation, equations of motion for the generalized metric contain contributions from the RR fields coming from the variation of the field $S$ that appears in the RR part of the action. The RR potentials of Type II theory in the democratic formulation are encoded in the Spin$(10,10)$ spinor~$\c$, such that the corresponding field strengths 
read
\begin{equation}
|F\rangle \equiv |\slashed\partial\chi\rangle=\sum_{p=1}^{10} \frac{1}{p!} F_{m_{1} \ldots m_{p}} \psi^{m_{1}} \cdots \psi^{m_{p}}|0\rangle, \quad \langle F| \equiv \langle\slashed\partial\chi|=\sum_{p=1}^{10} \frac{1}{p !}\langle 0| \psi_{m_{p}} \cdots \psi_{m_{1}} F_{m_{1} \ldots m_{p}}.
\end{equation}
Here the gamma matrices $(\y^m,\y_m)$ of Spin$(10,10)$ are defined in the usual way (up to rescaling)
\begin{equation}
    \{\y_m,\y^n\}= \d_m{}^n.
\end{equation}
To define Dirac conjugation one introduces the matrix
$A=\left(\y^{0}-\y_{0}\right)\left(\y^{1}-\y_{1}\right) \cdots\left(\y^{9}-\y_{9}\right)$:
\begin{equation}
\langle\overline{\slashed\partial\chi}|=\sum_{p=1}^{10} \frac{1}{p !}\langle 0|A\, \psi^{m_{p}} \cdots \psi^{m_{1}} F_{m_{1} \ldots m_{p}}.
\end{equation}
Finally, the kinetic operator $\mathcal{K}=A^{-1}S$ is written in terms of the Spin(10,10) image of the generalized metric, and it also contributes to the variation with respect to $\mH_{MN}$. 

The complete equations of motion of generalized metric of DFT for the action $\eqref{act}$ become (see \cite{Hohm:2011dv} for technical details)
\begin{equation}
\label{equa}
 e^{-2d} \mathcal{R}_{M N}+\mathcal{E}_{MN}=0,
\end{equation}
where the symmetric `stress-tensor' $\mE_{MN}$ with upper indices is defined as
\begin{equation}
\mathcal{E}^{M N}=\frac{1}{16} {\mathcal{H}_{P}{}^{(M}} \overline{\slashed{\partial} \chi} \Gamma^{N) P} \mathcal{K} \slashed \partial \chi=-\frac{1}{16} {\mathcal{H}_{P}{}^{(M}} \overline{\slashed{\partial} \chi} \Gamma^{N) P} \slashed \partial \chi.
\end{equation}{}
Explicitly in terms of the background RR field strengths this takes the following form
\begin{multline}
\label{RReqfil}
 \mathcal{E}_{MN}=-\frac{1}{8}\sqrt{-g}\left[\begin{matrix}2F_{m}F_{n}+F_{m pq}F_{n} {}^{pq}+\frac{1}{4!}F_{m pqrs}F_{n}{}^{pqrs}-g_{mn}\sum\limits_{i=1,3}|F^{(i)}|^2 \\
 2F^{n}{}_{m p}F^{p}+\frac{1}{3}F^{n}{}_{m pqr}F^{pqr}
 \end{matrix} \right.\\
 \left.\begin{matrix}
 2F^{m}{}_{n p}F^{p}+\frac{1}{3}F^{m}{}_{npqr}F^{pqr}\\
 2F^{m}F^{n}+F^{m}{}_{pq}F^{n pq}+\frac{1}{4!}F^{m}{}_{pqrs}F^{n pqrs}-g^{mn}\sum\limits_{i=1,3}|F^{(i)}|^2
 \end{matrix} \right],
 \end{multline}
where we see that the diagonal blocks are exactly the stress energy tensor for associated differential forms.

\subsection{Verifying the dilaton equation}

Consider now the dilaton equation of double field theory and perform non-abelian fermionic T-duality transformation, which maps $d \to d + \fr12 \log C$. The dilaton equation acquires additional terms 
\begin{equation}
\begin{aligned}
\D \mc{R} &= 2 \mathcal{H}^{M N} \partial_{M} \partial_{N} \log C- \mathcal{H}^{M N} \partial_{M} \log C \partial_{N}\log C \\
&-4\mathcal{H}^{M N} \partial_{M}d\, \partial_{N}\log C+2\partial_{M} \mathcal{H}^{M N} \partial_{N}\log C.
\end{aligned}
\end{equation}
The above must be zero for the dual background to solve the dilaton equation. Assuming that the initial background does do not depend on dual coordinates, we write the above explicitly as
\begin{equation}
    \begin{aligned}
       \D \mc{R} &= 2\, g^{mn}C^{-1}\dt_{m}\dt_{n}C - 3\, g^{mn}C^{-2}\dt_m C\, \dt_n C - 4\, g^{mn}C^{-1}\dt_m d\, \dt_n C + 2\, C^{-1}\dt_m g^{mn}\dt_n C\\
        &+4\, b^m{}_n C^{-1} \dt_m \tdt^n C - 6\, b^m{}_n C^{-2} \dt_m C\, \tdt^n C - 4\, b^m{}_n C^{-1}\dt_m d\, \tdt^n C + 2\, C^{-1}\dt_mb^m{}_n \tdt^nC\\
        &+2(g_{mn}-b_m{}^k b_{kn})C^{-1}\tdt^{m}\tdt^{n}C - 3(g_{mn} - b_m{}^kb_{kn})C^{-2}\tdt^m C\, \tdt^n C.
    \end{aligned}
\end{equation}  
Recall now the relation between the derivatives of the function $C=C(x^m,\tx_m)$ and the quadratic expressions built from the Killing spinors, which we denoted $K^m$ and $\tK_m$:
\begin{equation}
    \begin{aligned}
        \dt_m C& = iK_m - i b_{mn}\tK^n,\\
        \tdt^m C& = i\tK^m.
    \end{aligned}
\end{equation}
Substituting this into the expression for $\D \mc{R}$ one obtains
\begin{equation}
    \begin{aligned}
    \D \mc{R} &=  2 i g^{mn}C^{-1}\dt_{m}K_{n} + 3 g^{mn}C^{-2}K_m K_n  - 4i g^{mn}C^{-1}\dt_m d K_n + 2 C^{-1}i\dt_m g^{mn}K_n \\
    &+2 C^{-1} i b^m{}_n  \dt_m \tK^n  + 2(g_{mn}- b_m{}^k b_{kn})C^{-1}i\tdt^m \tK^n +3 C^{-2}g_{mn}\tK^m \tK^n.
    \end{aligned}
\end{equation}
One can check that the first line vanishes using the Fierz identities and the Killing spinor equations for $\e,\hat{\e}$. This is to be expected since the first line contains the new terms in the dilaton equation that appear when performing abelian fermionic T-duality. Terms in the second line are more subtle. Firstly, note that $\tdt^m \tK^n \equiv 0 $ since neither the initial background nor its Killing spinors depend on the dual coordinates. Similarly one can show $\dt_m \tK^n \equiv 0 $ :
\begin{equation}\label{constant-K-tilde}
    i \dt_m \tK^n = \dt_m \tdt^n C=\tdt^n \dt_m C=\tdt^n (K_m - b_{mk}\tK^k) \equiv 0.
\end{equation}
Finally, the last term $g_{mn}\tK^m \tK^n$ vanishes since it is proportional to terms of the form $\e \g^m \e\, \e \g_m \e$. 

It is noteworthy that our proposed prescription for non-abelian fermionic T-duality effectively restricts us to constant supersymmetric vector fields $\tilde{K}^m$ \eqref{constant-K-tilde}. This might be a hint that only abelian bosonic isometries can be accommodated by this formalism. However, examples of abelian fermionic T-duality are known, where the vector fields $K^m$ are non-constant~\cite{Bakhmatov:2009be}. Whether or not similar examples can be found for the vector fields $\tilde{K}^m$ in the non-abelian case is an open problem. In the present work we will only consider examples where there is a single (and constant) bosonic element $\tilde{K}^m$ in the subalgebra that we dualize. 

We conclude that the transformation~\eqref{fTtrans},~\eqref{eq:naftd} always generates backgrounds that solve the generalized dilaton equation of double field theory. This is true even if non-trivial dependence on the dual coordinates is generated. Although being a strong hint, this is not enough to claim that such transformation always gives solutions to all DFT equations, including the RR sector. To be on solid ground here one should additionally consider equations of motion for the generalised metric of DFT and for the RR O$(10,10)$ spinor. There is then a technical difficulty that while the transformation of the RR fields combined into an O$(1,9)$ bispinor are transparent, one has to translate them to the O$(10,10)$ language. One way of doing this is to decode explicitly the shifts of the RR field components from the shift of the bispinor by explicit contraction with gamma matrices. As usual, such a direct approach involves long and tedious calculations and requires dealing with both O$(1,9)$ gamma-matrices and O$(10,10)$ Clifford vacuum. Alternatively, a direct relationship between the bispinor and the DFT RR spinor components can be established. This depends on the particular embedding of O$(1,9)$ into O$(10,10)$ and on the component form of the Clifford vacuum state. Of help here can be the formalism developed in \cite{Jeon:2012kd}, where the RR fields are combined into an O$(1,9)$ bispinor rather than an O$(10,10)$ spinor.

Leaving these computations for future work, below we will provide a set of examples supporting the statement and establish a relation to the approach based on supercoset sigma models. Fermionic T-dual backgrounds in the examples also solve the Einstein DFT equation~\eqref{equa}, although this has not been checked in general. A proof that non-abelian fermionic T-dual backgrounds always solve the complete set of the double field theory equations deserves separate consideration in a future work.

\section{Sigma model perspective}
\label{sec:sigma}

The field theory description of non-abelian fermionic T-duality presented above together with examples of the next section provides strong evidence that it always yields a solution to double field theory equations of motion. It is natural to ask how string theory sigma model behaves under such transformation of its background fields. To see that we turn to the generic scheme of non-abelian duality transformations of the supercoset sigma model considered in~\cite{Borsato:2018idb}. Although the discussion there applies to generic isometry superalgebras, the authors concentrate on the bosonic case. Here we will briefly review the necessary results, then derive the non-abelian fermionic duality prescription using the formalism of~\cite{Borsato:2018idb} and check that it agrees with our proposal. 

While the results of~\cite{Borsato:2018idb} are valid for both supergroup and supercoset models, let us for simplicity restrict our consideration to the former. Then, denoting an element of the supergroup $g \in G$ and generators of the corresponding superalgebra as $T_{\mA}$, one defines the supervielbein $E^{\mA} = E_{\mM}{}^{\mA} dz^{\mM}$ in the usual way
\begin{equation}
    g^{-1} dg = E^{\mA} T_{\mA},
\end{equation}
where $z^{\mM}=(x^{m},\theta^\m,\hat\q^{\hat\m})$ are coordinates of the $d=10$ $\mathcal N=2$ superspace. The local frame index is split accordingly into bosonic and fermionic parts, $\mA = (a,\a,\hat\a)$. Lowest order $\q=\hat\q=0$ components of the supervielbein are related to the vielbein $e^{a}_{m}$ and the doublet of gravitini $\psi_{m}^{\alpha}, \psi_m^{\ah}$, however the most relevant for the discussion here is the spinor-spinor block
\begin{equation}
 \begin{pmatrix}
    E^\a_\m & E_\m^{\hat\a}\\
    E_{\hat\m}^\a & E^{\hat{\alpha}}_{\hat\m}
 \end{pmatrix}.
\end{equation}
It is always possible to represent the fermionic isometry that we are T-dualizing by the shift of a certain superspace fermion, $\q^1 \to \q^1 + \r$, where $\r$ is a constant Grassmann number. Then the above components of the supervielbein appear in the expressions $E_\mM^\a \d Z^\mM |_{\q=\hat\q=0} = \e^\a \r$ and $E_\mM^{\hat\a} \d Z^\mM |_{\q=\hat\q=0} = \e^{\hat\a} \r$, where we defined $\e^\a = E_1^\a |_{\q=\hat\q=0}$ and $\e^{\hat\a} = E_1^{\hat\a} |_{\q=\hat\q=0}$. Since $E_\mM^\a \d Z^\mM |_{\q=\hat\q=0}$ and $E_\mM^{\hat\a} \d Z^\mM |_{\q=\hat\q=0}$ in principle correspond to the local supersymmetry parameters, we conclude that $\e^\a$ and $\e^{\ah}$ are commuting Killing spinors of the supergravity background that is invariant under the shifts of $\q^1$.

We are now in a position to interpret the non-abelian T-duality rule for the RR bispinor $\mathcal{F}^{\a\hat\a}$ derived in~\cite{Borsato:2018idb}
\begin{equation}\label{borsato1}
\mc{F}'
= \hat{\Lambda} \Big( \mc{F}+16iE_{\mM} N^{\mM \mN}E_{\mN}|_{\q=\hat\q=0}\Big)
.
\end{equation}
In the second term one finds the matrix $N^{\mM \mN}$ whose inverse is defined by \begin{equation}\label{borsato2}
        N_{\mM \mN} = E_\mM{}^\mA E_\mN{}^\mB \left( G_{\mA \mB} - B_{\mA \mB} + \tilde{z}_{\mC} f_{\mA \mB}{}^\mC \right).
\end{equation}
Dual (super-)coordinates $\tilde{z}_{\mA}$ appear as Lagrange multipliers as in the standard T-duality procedure\footnote{For doubled superspace constructions see \cite{Hatsuda:2014qqa,Bandos:2015cha,Bandos:2016jez,Cederwall:2016ukd}}. The superalgebra structure constants $f_{\mA\mB}{}^\mC$ are defined as  usual in terms of the generators $T_\mA$. For the case of supergroups that we consider here,  $G_{\mA\mB}, B_{\mA\mB}$ are constant\footnote{They may depend on spectator fields in case when bosonic dimension of the supergroup is less than ten.} fields, which are tangent superspace representation of the Green-Schwarz superfields $G_{\mM\mN}$, $B_{\mM\mN}$, i.e.\
\begin{equation}
    \begin{aligned}
        G_{\mM\mN} & = E_{\mM}{}^{\mA}E_{\mN}{}^{\mB}G_{\mA\mB},\\
        B_{\mM\mN} & = E_{\mM}{}^{\mA}E_{\mN}{}^{\mB}B_{\mA\mB}.
    \end{aligned}
\end{equation}
Finally, the matrix $\hat{\L}$ in~\eqref{borsato1} is a $\mathrm{Spin}(1,9)$ transformation acting on one of the spinor indices, and corresponding to an O$(1,9)$ Lorentz transformation $\L$. These are defined as follows
\begin{equation}
    \begin{aligned}
        \L^a{}_b \G^b &= \hat{\L}^{-1}\G^a \hat{\L},\\
        \L^a{}_b & = \d^a{}_b - 2 E_{b \mM}N^{\mM \mN}E_{\mN}^a.
    \end{aligned}
\end{equation}
For a superalgebra consisting of a single Killing spinor, which anticommutes on a Killing vector $\tK^m$, the matrix $\hat{\L}$ corresponds to a single abelian T-duality along the Killing vector direction. Indeed, in this case the O$(1,9)$ transformation becomes simply an inversion
\begin{equation}
    \L^a{}_b = \d^a{}_b - 2 \tK^a \tK_b \tK^{-2},
\end{equation}
where $\tK^2 = G_{mn}\tK^m \tK^n$. This is precisely a single abelian T-duality along $\tK^m$ \cite{Sfetsos:2010uq}. Since we will be working in the DFT framework where abelian T-duality is an exchange of a coordinate for its dual, we assume $\hat{\L}=1$ for now and will take into account the bosonic T-duality at the very last step. 

It is easy to check that the above reproduces the standard abelian fermionic T-duality prescription. Assuming that the initial background has vanishing gravitino, only the spinor-spinor part of the supervielbein in~\eqref{borsato1} contributes:
\begin{equation}
    \d \mc F^{\a\ah} = 16\,i\, E_{1}^{\a}N^{11}E_{1}^{\hat{\a}}|_{\q=\hat\q=0} = 16\,i\, \e^\a \left(- B_{11}|_{\q=\hat\q=0} \right)^{-1} \eh^{\ah}.
\end{equation}
Note that the $1$'s above are fermionic indices. Dual coordinates $\tilde{z}^{\mM}$ do not appear due to the anticommutativity constraint $f_{11}{}^{\mA}=0$. Up to a conventional sign, this is precisely the abelian fermionic T-duality derived in~\cite{Berkovits:2008ic} with $C=B_{11}|_{\q=\hat\q=0}$. It is important to realize that $C=B_{11}|_{\q=\hat\q=0}$ is the component of the superfield $B_{\mM\mN}$ and may not be constant. Indeed, one can compute its derivative using the definition of the spinor-spinor-vector component of the superfield strength $H=dB$:
\begin{equation}\label{kuf}
    \begin{aligned}
        \left.\partial_{m}B_{11} \right|_{\theta=\hat{\theta}=0} = \left.\left( H_{m11} - 2\partial_{1} B_{m1} \right) \right|_{\q=\hat\q=0} = \left( E^{\mA}_{m} E^{\mB}_{1} E^{\mC}_{1} H_{\mA\mB\mC} - 2 \partial_1 \left( E^{\mA}_{1} B_{m\mA} \right) \right) \Big|_{\q=\hat\q=0}.
    \end{aligned}
\end{equation}
Using the type IIB supergravity constraints $H_{\alpha\beta c}=i(\bar\gamma_{c})_{\alpha\beta}$ and $H_{\hat{\alpha}\hat{\beta}c}=-i(\bar\gamma_{c})_{\hat{\alpha}\hat{\beta}}$, 
the first term above gives $ E^{\mA}_{m}E^{\mB}_{1}E^{\mC}_{1}H_{\mA\mB\mC} \big|_{\theta=\hat{\theta}=0} = i (\e\bar\g_m\e - \eh\bar\g_m\eh) = i K_m$\footnote{For type IIA, the constraint is $H_{\underline{\a} \underline{\b} c} = -i (\mathcal C \G_c \G_{11})_{\underline{\a} \underline{\b}}$ where the full $\underline\a=1,\ldots,32$ Dirac spinors and the corresponding gamma matrices are used~\cite{Howe:1983sra,Carr:1986tk,Wulff:2013kga}. Thus there is no relative sign in the definition of $K_m$ 
}. The second term may be nonzero due to the $\q$ dependence of $E_\m^a = \frac{i}{2} (\g^a)_{\r\m} \q^\r$:
\begin{equation}
    2 \partial_1 B_{m1} = 2 \partial_1 \left( E_1^a B_{ma} \right) = 2 \partial_1 \left( \frac{i}{2} (\g^a)_{\r 1} \theta^\r B_{ma} \right) = i (\g^n)_{11} B_{mn}.
\end{equation}
Finally,~\eqref{kuf} becomes
\begin{equation}\label{lamed}
    \left.\partial_{m}B_{11} \right|_{\theta=\hat{\theta}=0} = i K_m - i b_{mn} \tilde{K}^n,
\end{equation}
where the Kalb-Ramond field $b_{mn}$ is the lowest order component field of $B_{mn}$, and we have converted the world fermionic indices on the gamma matrix to tangent space with the help of the Killing spinors, $(\g^n)_{11}|_{\theta=\hat{\theta}=0} = \e^\a (\g^n)_{\a\b} \e^\b + \eh^{\hat\a} (\g^n)_{\hat\a\hat\b} \eh^{\hat\b} = \tilde{K}^n$.
For the abelian fermionic isometry the second term does not contribute since $\tK^m=0$ is the abelian constraint. Thus, we recover the abelian fermionic T-duality prescription~\cite{Berkovits:2008ic}.

Moving on, let us consider the simplest case of non-abelian duality with respect to one superspace fermionic shift only. This comes about with just a pair of Killing spinors $\e,\eh$ that violate the anticommutativity relation, $\tK\neq 0$, so that both terms in~\eqref{lamed} contribute. We should also take into account the non-vanishing structure constant $f_{11}{}^m =i\tK^m \neq 0$ in~\eqref{borsato2}. The transformation of the RR fields still can be written as $\d \mc F^{\a\ah} = 16\,i\, \e^\a C^{-1} \eh^{\ah}$, if we redefine $C$ to take care of the new terms:
\begin{equation}\label{x-tilde}
    C = (B_{11} + \tilde z_{\mM}f_{11}{}^{\mM}) |_{\theta=\hat{\theta}=0}  = B_{11} |_{\theta=\hat{\theta}=0} + i \tilde{x}_{m}\tilde{K}^{m},
\end{equation}
where $B_{11}|_{\theta=\hat{\theta}=0}$ satisfies~\eqref{lamed}. One immediately infers the dual coordinate derivative $\tdt^m C = i \tK^m$. This can be viewed as one of the defining equations for $C$ in the non-abelian case when starting from the Killing spinors. This equation holds beyond supercoset backgrounds as we have shown above. 

From \eqref{x-tilde} it is clear that an extra bosonic T-duality in the direction of $\tilde K^m$ will strip the background of any dual coordinate dependence, as $\tilde x_m$ becomes geometric after the T-duality. This is a robust means of ending up in standard supergravity, rather than DFT. A necessary condition is that the first term of~\eqref{x-tilde} does not depend on $x^m$, which would become non-geometric after the T-duality. 

In the second defining equation for $\dt_m C$ we have the standard contribution $i K_m$ plus extra terms:
\begin{equation}
\partial_{m}C = i K_{m} - i b_{mn} \tilde{K}^{n} + i \tx_n \dt_m \tK^n.
\end{equation}
We have seen in the previous section that the derivative $\dt_m C$ defined as above, but without the last term, ensures that the generalized dilaton equation of DFT holds. With the extra term $i \tx_n \dt_m \tK^n$ that equation gets new contributions of different orders in $\tx_m$, contracted with expressions that depend purely on geometric coordinates. Hence, the only condition for this to satisfy the generalized dilaton equation of motion is
\begin{equation}
    \label{eq:extracond}
    \tx_n \dt_m \tK^n = 0
\end{equation}
and we recover precisely the same defining PDE's for $C$ as in \eqref{eq:naftd}.

At this point the origin of the above condition is not completely clear and more detailed analysis of the sigma model is needed. From the field theory point of view this is simply imposed by the equations of motion. In section~\ref{sec:examples} we will see that this condition indeed is satisfied for the non-trivial example of non-abelian fermionic T-duality of the D-brane background. We conclude that the field transformations~\eqref{fTtrans} with $C$ defined by~\eqref{eq:naftd} for general supersymmetric backgrounds agree with the results of~\cite{Borsato:2018idb} in the case of supercoset sigma models.

\section{Examples}
\label{sec:examples}

\subsection{The Minkowski space}

As a proof-of-concept and an elementary example let us consider the flat empty spacetime in $d=10$, which is a maximally supersymmetric solution, i.e.\ one has 32 fermionic directions given by the degenerate doublets of constant Killing spinors $(\e_i,0)$, $(0,\eh_i)$, $i\in \{1,\ldots,16\}$:
\begin{equation}
\begin{aligned}
(\e_i)^\a=\d_{i}{}^\a, \quad (\eh_{j})^{\ah}=\d_{j}{}^{\ah}.
\end{aligned}
\end{equation}
Since the space-time metric $g_{mn}$ is constant and the $b$-field is zero, the generalized Ricci tensor can be written completely in terms of derivatives of the generalized dilaton
\begin{equation}
\mathcal{R}_{M N}=\begin{pmatrix}  \partial_{m} \partial_{n}-g_{ml}g_{nk}\tilde{\partial}^{l}\tilde{\partial}^{k} &g^{ml}g_{nk}\partial_{l} \tilde{\partial}^{k}-\tilde{\partial}^{m}\partial_{n}\\
g_{ml}g^{nk}\tilde{\partial}^{l} \partial_{k}-\partial_{m}\tilde{\partial}^{n}&\tilde{\partial}^{m} \tilde{\partial}^{n}-g^{ml}g^{nk}\partial_{l}\partial_{k}\end{pmatrix}d.
\end{equation}
This means that the generalized Einstein equations~\eqref{equa} simplify a lot. Below we provide several examples of non-abelian fermionic T-duality of Minkowski spacetime based on various choices of the Killing spinors. These are selected so as to highlight certain generic properties of non-abelian fermionic T-dual backgrounds. All the examples solve the field equations of the NSNS sector of double field theory.

{\bf Example 1.} The spinors are taken to be
\begin{equation}
\label{Kil135}
(\e_1 - i \e_9, -\eh_1 - i \eh_9).
\end{equation}
Using the explicit realization of the gamma matrices, one can check that this direction is non-abelian, and the doubled spacetime derivatives of $C$~\eqref{eq:naftd} are $\dt_8C = 4$, $\tdt^9 C = 4i$, leading to
\begin{equation}\label{C}
C = 4 (x^8 + i\tx_9).
\end{equation}
One may recover the fermionic T-dual background using~\eqref{fTtrans}. The metric is still flat and no $b$ field is generated. We do get a nontrivial dilaton
\begin{equation}\label{dual1}
    \phi = \frac12 \log 4(x^8 + i \tx_9),
\end{equation}
and the RR fluxes (assuming that we are in type IIB):
\begin{equation}
\begin{aligned}\label{dual2}
    F_0 &= -2\, i\, C^{-3/2},\\
    F_{089} &= F_{127} = -F_{134} = -F_{156} = F_{235} = -F_{246} = F_{367} = F_{457} = -2\, C^{-3/2},\\
    F_{01236} &= F_{01245} = -F_{01357} = F_{01467} = -F_{02347} = -F_{02567} = F_{03456}= \\
    F_{12789}&=- F_{13489} = -F_{15689}= F_{23589} = -F_{24689} = F_{36789} = F_{45789}= 2\,i\, C^{-3/2}.
\end{aligned}
\end{equation}
Overall this looks very reminiscent of abelian fermionic T-duality transformation. The main difference is of course that this background is not a supergravity solution due to the explicit dual coordinate dependence of $C$. However, it can be shown to satisfy equations of motion of double field theory. The easiest way to do this is to perform a bosonic T-duality along $x^9$, which is equivalent to replacing $\tx_9 \leftrightarrow x^9$ in all expressions. This will bring us to the Type IIA theory, still with flat metric and no $b$ field, the dilaton $\phi = \frac12 \log 4 (x^8 + i x^9)$ and the RR fluxes reshuffled according to $F'^\a{}_{\hat\b} = F^{\a\hat\d} (\bar\g^9)_{\hat\d\hat\b}$. Explicit computation shows that supergravity field equations are then satisfied. This implies that~\eqref{dual1},~\eqref{dual2} is a complex-valued solution to the DFT equations of motion, which one can also verify directly, using the equations of section~\ref{sec:dft}. It belongs to the geometric orbit of T-duality, i.e.\ one bosonic T-duality takes it to a (complex-valued) solution of supergravity.

{\bf Example 2.} One may consider a fermionic T-duality transformation generated simply by a single Majorana spinor of flat spacetime, which is also non-abelian:
\begin{equation}
(\e_1, 0).
\end{equation}
Then the fermionic T-dual is Minkowski with non-trivial doubled spacetime dependence of the dilaton,
\begin{equation}\label{dual}
    \begin{aligned}
        g_{mn}&=\h_{mn}, \quad b_{mn}=0,\\
       e^{2\phi}&=i(x^0-\tx_0+x^9+\tx_9),\\
       F_{RR}&=0,
    \end{aligned}
\end{equation}
Despite the exotic dependence on dual coordinates via combinations of the type $x\pm \tx$, the section constraint is satisfied. No RR flux is generated in this transformation, because we have chosen $\eh=0$.

The background can be made real valued by adjusting the phase of the Killing spinor,
\begin{equation}
    (e^{i\fr \p 4}\e_1, 0),
\end{equation}
which results in the dilaton $e^{2\phi}= -x^0+\tx_0-x^9-\tx_9$. This real background is too a solution of double field theory, however it is still not completely satisfactory as it includes dependence on the dual time. In fact, it is impossible to avoid dependence on the dual time without using linear combinations of Killing spinors with relative complex phase shifts between different terms. The reason is that the supersymmetric Killing vector $\tK^m$ is always timelike or null in real valued supergravity~\cite{Gibbons:1982fy,Tod:1983pm}, which implies $\tx_0$ dependence by virtue of~\eqref{eq:naftd}. 

A simple check shows that the above background satisfies the DFT field equations of section~\ref{sec:dft}, and hence the corresponding fermionic T-duality produces a genuinely non-geometric solution. Let us elaborate more on what we understand by genuine non-geometry here. Although the background has non-trivial dependence on the dual coordinates $\tx_0,\tx_9$, this can be removed by an appropriate O$(2,2)$ transformation, since the section constraint is satisfied \cite{Hohm:2010jy}. Indeed, consider combinations $x^\pm = 1/\sqrt{2}(x^0 \pm x^9)$ and $\tx_\pm = 1/\sqrt{2}(\tx_0 \pm \tx_9) $, which are related to the initial coordinate basis by an O$(2,2)$ rotation. The background~\eqref{dual} takes the form
\begin{equation}
    \begin{aligned}
    g_{mn} & = 
    \begin{pmatrix}
         \begin{matrix}
         0 & -1  \\
         -1 & 0  
         \end{matrix} & \mathbf{0}_{2\times 8} \\
         \mathbf{0}_{8\times 2} & \mathbf{1}_{8\times 8}
    \end{pmatrix}, \quad b_{mn} = 0\\    
    e^{2\phi} & = i\sqrt{2}\, (x^+ - \tx_-), \\
    F_{RR} & = 0.
    \end{aligned}
\end{equation}
Since the invariant metric $\h_{MN}$ is preserved and has the same form in the new basis, the coordinates $x^\pm$ and $\tx_\pm$ are mutually dual, and a further T-duality along $x^-$ is possible. This removes the dependence on $\tx_-$ rendering the dilaton (in the old coordinates) $e^{2\phi} = 2 i x^9$. However, the lower right block of the generalised metric $\mH^{mn}$ becomes degenerate, which means that the Riemannian metric cannot be  defined (at least not in the B-frame). Still being a solution to DFT equations of motion such string backgrounds have been referred to as non-Riemannian in \cite{Lee:2013hma} (see \cite{Morand:2017fnv} for full the classification).

In \cite{Dibitetto:2012rk} non-geometric string backgrounds have been considered in the context of generalised Scherk-Schwarz compactifications of double field theory and of the scalar sector of exceptional field theories. The corresponding flux configurations can be naturally divided into two large classes: geometric and genuinely non-geometric. The former correspond to such fluxes, both geometric or not, that can be T-dualised into a geometric flux configuration. In other words, those belonging to a T-duality orbit that contains a geometric configuration. It is shown, that in this case the configuration can be uplifted to the maximal supergravity. In contrast, an orbit of a genuinely non-geometric flux configuration does not contain a geometric representative and hence cannot be uplifted to the maximal supergravity. Such flux configurations are generated by generalised twist matrices that break the section constraint. Hence, while the lower dimensional geometry is well defined, one is not able to present a full 10-dimensional geometric description.

Something similar we observe in the example above. One is either in the non-geometric frame, i.e.\ one has to deal with a background that depends on dual coordinates, or is in the frame where no Riemannian metric can be defined. Although potentially a matter of debate, we think it natural to also classify such backgrounds as genuinely non-geometric.

{\bf Example 3.} An example of a combination that provides more sensible, however still complex background is:
\begin{equation}
     \begin{aligned}
     \e&=\fr1{\sqrt{2}}(\e_1+i\e_9),\\
     \eh&=0.
    \end{aligned}
\end{equation}
This leads to the following genuinely non-geometric solution
\begin{equation}
    \begin{aligned}
        g_{mn}&=\h_{mn}, \quad b_{mn}=0,\\
       e^{2\phi}&=-x^8-\tx_8+i(x^9+\tx_9),\\
       F_{(p)}&=0.
    \end{aligned}
\end{equation}
Hence, non-abelian fermionic T-duality gives a genuinely non-geometric solution of DFT, which is complex and does not depend on dual time.

\subsection{D{\it p}-brane backgrounds}\label{sec:Dbrane}

Moving on to less trivial examples, we can consider D$p$-brane solutions in $d=10$:
\begin{equation}
\label{Dmetr}
\begin{gathered}
ds^2_{(p)} = H_{(p)}^{-\fr12} \left[ -dt^2 + dy_{(p)}^2 \right] + H_{(p)}^{\fr12} dx_{(9-p)}^2,\\
e^{-2(\phi-\phi_0)} = H_{(p)}^{\frac{p-3}{2}},\quad A = e^{-\phi_0} (H_{(p)}^{-1} - 1)\, dt\wedge dy_1\wedge\ldots\wedge dy_{p},
\end{gathered}
\end{equation}
where $H_{(p)}$ is a function of the transverse directions $x_{(9-p)}$, and is a harmonic function in a given dimension. The solutions preserve half of the maximum supersymmetry, due to the BPS constraints being the projection conditions 
 \begin{equation}
  \left( 1 - \Gamma^{0\ldots p} O_{p} \right) 
  {\e \choose \eh}= 0,
 \end{equation}
where $\mathcal{O}_{p}$ depends on the dimension as well as on type IIA versus type IIB theory and is such as to make the above operator a projector. Overall, there are 16 independent Killing spinors in the D$p$-brane background, parametrized by an arbitrary constant Majorana-Weyl spinor $\eh_0$ of appropriate chirality:
\begin{equation}\label{D3spinor}
\begin{aligned}
 \e &= 
    \left\{
        \begin{aligned}
            &H_{(p)}^{-\fr18} i \g^{0\bar 1\ldots p} \eh_0 \qquad &(p \mathrm{~even,~IIA),}\\ 
            &H_{(p)}^{-\fr18} \g^{0\bar 1\ldots \bar p} \eh_0 \qquad &(p \mathrm{~odd,~IIB),}
        \end{aligned}
    \right.\\
 \eh &= H_{(p)}^{-\fr18} \eh_0.
\end{aligned}
\end{equation}

Expressions for the Killing spinor bilinears $K_m, \tK^m$ simplify substantially due to the gamma matrix identities 
\begin{align}
    (\g^{0\bar 1\ldots p})^T \bar\g^a (\g^{0\bar 1\ldots p}) &= 
        \left\{
            \begin{aligned}
                -&\g^a,\quad a\leq p,\\
                &\g^a,\quad a > p,
            \end{aligned}        
        \right.\qquad(p \mathrm{~even,~IIA),}\\
    (\g^{0\bar 1\ldots \bar p})^T \bar\g^a (\g^{0\bar 1\ldots \bar p}) &= 
        \left\{
            \begin{aligned}
                &\bar\g^a,\quad a\leq p,\\
                -&\bar\g^a,\quad a > p,
            \end{aligned}        
        \right.\qquad(p \mathrm{~odd,~IIB).}
\end{align}
Recalling that we have alternating signs in the type IIA versus type IIB expressions for $K_m$~\eqref{Cdef} and $\tK^m$~\eqref{constr}, we end up with the same answer for any D$p$-brane,
\begin{align}
\label{partC}
    \left.
    \begin{aligned}
        \dt_m C &= 0\\ 
        \tilde\dt^m C &= 2i\,\d_a^m \eh_0 \g^a \eh_0
    \end{aligned}
    \;\right\}
    \quad m \leq p,\\
    \left.
    \begin{aligned}
        \dt_m C &= 2i\,\d^a_m \eh_0 \g_a \eh_0 \\
        \tilde\dt^m C &= 0 
    \end{aligned}
    \;\right\}
    \quad m > p,
\end{align}
(replace $\g_a \to \bar\g_a$ according to the chirality of $\eh_0$). It is noteworthy that the vielbein and the metric have worked out precisely so that the right hand sides above are constant, as expected~\eqref{constant-K-tilde}.

We observe that the function $C$ cannot depend on coordinates dual to the transverse directions. It may depend on coordinates dual to the isometric directions along the worldvolume. Hence, the combination $\tx_m \tK^m$ is non-zero and in fact does not depend on geometric (transverse, in this case) coordinates, thus satisfying the condition~\eqref{eq:extracond}. Also recall that we had the condition $\partial_{m}g \, \tilde{\partial}^m C=0$ as a consequence of the strong constraint in DFT~\eqref{strong}, and this is now satisfied identically.

As a specific example let us consider fermionic T-duality for the D3-brane. We can choose the arbitrary constant spinor in~\eqref{D3spinor},
\begin{equation}
    \eh_0^\a = \frac{1}{2\sqrt{2}}e^{\frac{i\pi}{4}} (-\delta_1^\a + i \delta_2^\a + \delta_{15}^\a + i \delta_{16}^\a),
\end{equation}
so that it yields a simple value for the duality parameter:
\begin{equation}
    C = x^4 + i \tx_1.
\end{equation}
This appears in the dual dilaton $e^{2\phi} = e^{2\phi_0} C$, as well as in the RR fields
\begin{align}
    &F_{(1)} = -\frac{e^{-\phi_0}}{2 C^{3/2}} \, dx^6, \\
    &\begin{aligned}
        F_{(3)} = \frac{i e^{-\phi_0}}{2 C^{3/2}} &\left[ dx^0 \left(H^{-1} dx^{23} +dx^{58} -dx^{79}\right) - dx^{146}\right.,\\
        &+ i dx^2 (dx^{57} + dx^{89}) +  i dx^3 (dx^{59} + dx^{78})\big],
    \end{aligned}\\
    &\begin{aligned}
        F_{(5)} = -\frac{e^{-\phi_0}}{2 C^{3/2}} & \left[ \sum_{k=4}^9 H^{-1}\left( \delta_k^4 + 2 C H^{-1} \dt_k H  \right) dx^{0123k} + dx^{014} (dx^{58} - dx^{79})\right. \\
        &- i dx^{06} \left( dx^2 (dx^{59} + dx^{78}) + dx^3 (dx^{57} + dx^{89})\right)\bigg].
    \end{aligned}
\end{align}
We have employed the obvious notation $dx^{mn} = dx^m\wedge dx^n$, etc., in order to keep the expressions compact. This background is a DFT solution and can be mapped to a solution of supergravity by a bosonic T-duality in $x^1$.

\section{Conclusions}\label{sec:conc}	

To summarize, we have proposed and worked out a supergravity description of a non-abelian generalization of fermionic T-duality along a subalgebra of the full supersymmetry algebra consisting of a Killing spinor $(\e,\hat{\e})$ and the corresponding Killing vector $\tK^m$. The construction hinges on the observation that relaxing the abelian constraint $\tK^m=0$ for the Killing spinors may be interpreted as making the fermionic T-dual background depend on the dual coordinates of double field theory. Within the DFT framework, the section constraint that ensures consistency of a field configuration is automatically satisfied. In the specific case of supercoset backgrounds our results agree with those obtained for the Green-Schwarz superstring on coset spaces~\cite{Borsato:2018idb}. However, on top of the sigma model description we get an additional constraint $\tx_m\dt_n \tK^m=0$, where $\tx_m$ are the dual (winding) coordinates. From the field theory point of view, adding such a term to the scalar function $C$ as prescribed by the sigma model procedure violates equations of motion of DFT. Detailed analysis of the sigma model origin of this constraint is beyond the scope of the present paper. 

We have seen that non-abelian fermionic T-duality generates solutions of double field theory equations of motion. Checking this for the explicit examples: empty Minkowski spacetime and D$p$-brane backgrounds, we encounter several different cases summarized in Table \ref{tab:cases}.
\begin{table}[ht]
    \centering
    \begin{tabular}{|c|c|c|}
            \hline
          Fields & Dependence on  the dual time & Orbit\\
         \hline 
          Complex & no & geometric \\
          Real & yes & non-geometric\\
          Complex & no & non-geometric\\
         \hline
    \end{tabular}
    \caption{Typical backgrounds generated by a non-abelian fermionic T-duality.}
    \label{tab:cases}
\end{table}
We observe that at least for the spacetimes that we consider it is impossible to generate a real valued solution with no dual time dependence. The reason is the fundamental fact that $\tilde K^2 \geq 0$ in real valued supergravity~\cite{Gibbons:1982fy,Tod:1983pm}. Dependence on the dual time makes the background effectively complex due to the wrong sign of the kinetic terms of the RR fields. 

An independent source of complex values in fermionic T-duality is an explicit factor of $i$ in the last term of the transformation of the RR fields~\eqref{fTtrans}. Analysis shows that when dualizing along a single fermionic isometry, the interplay of the two aforementioned factors in general leads to complex valued dual backgrounds. It is likely that promoting non-abelian fermionic T-duality to the case of more than one non-commuting Killing spinor direction may help to develop an algorithm for generating real valued supergravity solutions. 

Apart from the reality property of the dual backgrounds, of interest is their dependence on dual coordinates reflected by the last column of Table~\ref{tab:cases}. One typically finds $C$ to be a linear function of the form $\alpha x^m + \beta \tx_n$, where $x^m$ and $\tx_n$ are some particular geometric and dual coordinates. In the case $m\neq n$ the background belongs to the geometric orbit of bosonic T-duality in the sense of~\cite{Dibitetto:2012rk}. Indeed, then a T-duality along $\tx_n$ would turn it into a geometric $x^n$ and all dual coordinate dependence is gone. This corresponds to T-dualizing a closed sub-algebra consisting of a non-abelian fermionic generator and a bosonic generator $\tilde K^m$. Otherwise, when $m=n$ removing the dual coordinate dependence by means of a bosonic T-duality may lead to a non-Riemannian background, as was demonstrated in the examples section. Hence the corresponding background belongs to a non-geometric T-duality orbit and is referred to as genuinely non-geometric. Examples of such genuinely non-geometric backgrounds have been found in~\cite{Dibitetto:2012rk} in generalized Scherk-Schwarz reductions of DFT for twist matrices violating the section constraint. In contrast, genuinely non-geometric backgrounds of the form we obtain respect the section constraint and are valid string theory backgrounds.

For some examples that we consider one finds vanishing RR fields in the dual background and linear dependence on the dual coordinate in the scalar function $C$. Although, the dilaton is related to $C$ via the exponent, this reminds the double and exceptional field theory description~\cite{Sakatani:2016fvh,Baguet:2016prz} of generalized supergravity~\cite{Arutyunov:2015mqj}. There one considers a solution of the section constraint, where the metric and the background gauge fields depend on at most nine spacetime coordinates, while the dilaton is allowed to depend linearly on the remaining tenth coordinate $x^9$. Then bosonic T-duality along $x^9$ turns it into $\tx_9$ and introduces  linear dependence on the dual coordinate. For the Green-Schwarz superstring one can check that such backgrounds are allowed by the kappa-symmetry~\cite{Sakamoto:2017wor,Fernandez-Melgarejo:2018wpg}. Thus it may be interesting to inspect string behaviour on backgrounds produced by non-abelian fermionic T-duality. 

Another interesting further direction is to investigate possible relations between non-abelian fermionic T-dualities and deformations of supercoset sigma models corresponding to Dynkin diagrams with all fermionic simple roots. Bivector deformations of integrable sigma models on supercosets are known to correspond to bosonic non-abelian T-dualities, at least for a certain class where the $r$-matrix is invertible~\cite{Borsato:2017qsx,Borsato:2018idb}.  Integrable $\h$-deformations of 10-dimensional backgrounds with an AdS factor have been found to produce solutions of the ordinary supergravity equations when the deformation is done along fermionic simple roots of the isometry superalgebra~\cite{Hoare:2018ngg} (see~\cite{Seibold:2020ouf} for a review). 

Finally, of particular interest is the admittedly long-standing problem of self-duality of the AdS${}_4\times \CP^3$ background. So far searches for a chain of fermionic and bosonic T-duality that map the background to itself have all failed~\cite{Adam:2010hh,Bakhmatov:2010fp,Colgain:2016gdj}, and one may hope that the novel non-abelian fermionic T-duality might be helpful here. Indeed, conventionally one would try to organise the chain such that each T-duality maps a solution to a solution of supergravity equations of motion. As we have seen, this greatly restricts the possible choices of the Killing spinors, but there is no such restriction in the non-abelian case. Since solutions of DFT equations of motion can be understood as proper background of the double or even the standard superstring, one could hope to organise such a chain which goes beyond the set of supergravity backgrounds and maps AdS${}_4\times \CP^3$ to itself.

\section*{Acknowledgements}

This work has been supported by Russian Science Foundation under the grant agreement RSCF-20-72-10144. The work of IB covered in section~\ref{sec:Dbrane} was done with the partial support from the Russian Government program of competitive growth of Kazan Federal University.

\appendix
\section{Index summary and spinor conventions}
\label{app:spinor}

Some of the indices that we use are:
\begin{align*}
&& m,n;\, a,b &\in \{0, \ldots, 9\} && \mathrm{Lorentz~vector~index~(world/tangent)},&&\\
&& \m,\n;\,\a,\b &\in \{1, \ldots, 16\} && \mathrm{Weyl~spinor~index~(world/tangent)},&&\\
&& M,N &\in \{0, \ldots, 19\} && \mathrm{double~spacetime~coordinate~index}.
\end{align*}
Hatted spinor indices $\ah,\bh$ correspond to the second spinor of $\mathcal{N}=2$ supersymmetry. It can be opposite (type IIA) or same (type IIB) chirality, compared to the first spinor with unhatted index:
\begin{align}
    \text{IIA:}\qquad 
                \e = \begin{bmatrix}
                            \e^{\a} \\ 0 
                       \end{bmatrix},
                \quad 
                \eh = \begin{bmatrix}
                            0 \\ \eh_{\ah} 
                       \end{bmatrix}, \\
    \text{IIB:}\qquad 
                \e = \begin{bmatrix}
                            \e^{\a} \\ 0 
                       \end{bmatrix},
                \quad 
               \eh = \begin{bmatrix}
                           \eh^{\ah} \\ 0
                       \end{bmatrix}.
\end{align}
We are using a Majorana-Weyl representation of the $SO(1,9)$ gamma  matrices, such that
\begin{equation}
    \G^m
    =
    \begin{bmatrix}
        0 & (\g^m)^{\a\b} \\
        (\bar\g^m)_{\a\b} & 0
    \end{bmatrix},\qquad
    \mathcal{C}
    =
    \begin{bmatrix}
        0 & c_\a{}^\b \\
        \bar c^\a{}_\b & 0
    \end{bmatrix},\qquad
    \psi
    = 
    \begin{bmatrix}
        \phi^\a \\
        \chi_\a
    \end{bmatrix}.
\end{equation}
The charge conjugation matrix can be used to define Majorana conjugation, relating covariant and contravariant spinors, $\bar\psi = \psi^T \mathcal C$. One can look up an explicit example of the gamma matrices in this representation e.g.\ in~\cite{Bakhmatov:2009be}. The important algebraic properties of the $\g$ matrices are $(\g^m)^{\a\b} (\bar\g^n)_{\b\g} + (\g^n)^{\a\b} (\bar\g^m)_{\b\g} = 2 \h^{mn} \delta^\a_\g$ and $(\g^m)^{(\a\b} \g_m{}^{\g)\d}=0$. Both $\g$ and $\bar\g$ are symmetric, as are $\g^{m_1\ldots m_5}$; $\g^{m_1m_2m_3}$ on the contrary are antisymmetric.  In the main text we sometimes use the obvious notation $\g^{a\bar b \ldots c} = \g^{[a}\bar\g^{b}\ldots \g^{c]}$.

Depending on chirality, one gets for the commutator of the supersymmetry transformation with itself
\begin{equation}
     [\bar\e Q, \bar\e Q] = 
     \left\{
        \begin{aligned}
            - &(\e c)^\a \left\{ Q_\a, Q_\b \right\} (\e c)^\b,\qquad \textrm{for}\:\, \e = \begin{bmatrix} \e^\a \\ 0 \end{bmatrix},\\
            - &(\e \bar c)_\a \left\{ Q^\a, Q^\b \right\} (\e \bar c)_\b,\qquad \textrm{for}\:\, \e = \begin{bmatrix} 0 \\ \e_\a \end{bmatrix}.
        \end{aligned}
     \right.
\end{equation}
Thus, for $\mathcal{N}=2$ $d=10$ supersymmetry with two independent charges $Q, \hat Q$, employing the fundamental relations of the supersymmetry algebra, $\{Q, Q \} = (\mathcal C\G^m) P_m = \{\hat Q, \hat Q \}$, we have
\begin{equation}
    [\d_{\e,\eh}, \d_{\e,\eh}] = [\bar\e Q, \bar\e Q] + [\bar\eh \hat Q, \bar\eh \hat Q] = 
    \left\{
        \begin{aligned}
            -&(\e \bar\g^m \e - \eh \g^m \hat \e) P_m,\quad \text{IIA},\\
             -&(\e \bar\g^m \e + \eh \bar\g^m \eh) P_m,\quad \text{IIB},
        \end{aligned}    
    \right.
\end{equation}
where we restrict to a gamma matrix representation with $c_\a{}^\b = \d_\a^\b$, $\bar c^\a{}_\b = -\d^\a_\b$. This motivates the definition of the vector field $\tilde K^m$~\eqref{constr}. 
Vanishing of $\tK^m$ is simply a criterion that the supersymmetry which corresponds to the fermionic shift $\d_{\e,\eh}$ is abelian.

Supersymmetry transformations for type IIB are given by
\begin{equation}
\label{susy-vars}
\begin{aligned}
\d\psi_m &= \nabla_m\e - \frac14 \slashed{H}_m\e - \frac{e^\f}{8} \left( \slashed{F}_{(1)} + \slashed{F}_{(3)} + \frac12 \slashed{F}_{(5)} \right) \bar\g_m \eh \\
\d\hat\psi_m &= \nabla_m\eh + \frac14 \slashed{H}_m \eh + \frac{e^\f}{8} \left( \slashed{F}_{(1)} - \slashed{F}_{(3)} + \frac12 \slashed{F}_{(5)} \right) \bar\g_m \e,\\
\d\l &= \slashed{\dt}\f\,\e -\frac12 \slashed{H} \e + \frac{e^\f}{2} \left( 2\slashed{F}_{(1)} + \slashed{F}_{(3)} \right) \eh,\\
\d\hat\l &= \slashed{\dt}\f\,\eh +\frac12 \slashed{H} \eh - \frac{e^\f}{2} \left( 2\slashed{F}_{(1)} - \slashed{F}_{(3)} \right) \e,
\end{aligned}
\end{equation}
where
\begin{align}
\slashed{F}_{(n)} = \frac{1}{n!} F_{m_1\ldots m_n} \g^{m_1\ldots m_n},\qquad
\slashed{H}_m = \frac12 H_{mnk} \g^{nk}.
\end{align}

\bibliographystyle{JHEP}
\bibliography{bib.bib}
\end{document}